\newif\iftwocolumn \twocolumnfalse
\def\Vec#1{{\bf #1}}
\def\RE{{\rm Re}}
\def\IM{{\rm Im}}
\newcommand{\posfig}{t}
\twocolumntrue    
\iftwocolumn
\documentstyle[floats,twocolumn,aps,prb,epsf]{revtex}  
\else
\documentstyle[prb,aps,manuscript,epsf]{revtex}
\fi
\begin{document}
\draft
\title{
Systematic Design of Antireflection Coating for
 Semi-infinite One-dimensional Photonic Crystals Using Bloch Wave Expansion
}%
\author{Jun Ushida, Masatoshi Tokushima, Masayuki Shirane, and Hirohito Yamada}
\address{Fundamental Research Laboratories, NEC Corporation,
 34 Miyukigaoka, Tsukuba, 305-8501, JAPAN}
\iftwocolumn
\else
\date{\today} 
\fi
\iftwocolumn
\twocolumn [ 
\fi
\maketitle
\begin{abstract}
\iftwocolumn
\hspace*{1.5cm}\parbox{15cm}{  
\fi
We present a systematic method for designing a perfect 
antireflection coating (ARC) for a semi-infinite one-dimensional (1D) 
photonic crystal (PC) with an arbitrary unit cell.
We use Bloch wave expansion and time reversal symmetry, 
which leads exactly to analytic formulas of structural parameters 
for the ARC and renormalized Fresnel coefficients of the PC.
Surface immittance (admittance and impedance) matching plays an essential role 
in designing the ARCs of 1D PCs, which is shown together 
with a practical example. 
\iftwocolumn
 } 
\fi
\end{abstract}
\iftwocolumn
 \hspace*{1.5cm}\parbox{15cm}{ 

\fi
\iftwocolumn
\pacs{$Revision: 4.16 $\hspace*{1cm}$Date: 2003-01-06 08:54:04+09 $}
\else
\fi
\iftwocolumn
 } 
 ] \narrowtext 
\fi
Photonic crystals (PCs) have unique energy dispersion 
due to coupling between periodic materials and electromagnetic (EM) waves.
\cite{Yab,Jo,Ohtaka,Sakoda}
In particular, the strong energy dispersion of the propagation modes of PCs
has attracted much interest because it enables applying PCs to
add/drop multiplexers, dispersion compensators, polarization filters,
and image processors.\cite{Kosaka,Kawakami,Notomi}
These transmission-type applications require a 
negligibly small reflection loss at the PC interface.\cite{Baba} 
Therefore, applying effective antireflection interface structures, 
or antireflection coatings (ARCs) to the input and output 
interfaces of the PCs is important.\cite{Baba,Xu,Mekis2,Mekis3}
Several ARCs designed for the PCs have been reported on.
The structural ideas of these ARCs were based on the 
concepts of adiabatic interconnection\cite{Baba,Xu,Mekis2} 
or wave vector matching ($\Vec{k}$ matching)\cite{Mekis2,Mekis3} for 
two-dimensional PCs.
The reflectance at the interface of one-dimensional (1D) PCs has 
so far been calculated by
plane-wave expansion and multiplication of the transfer matrix for 
a unit cell.\cite{Yeh2,BornWolf,Macleod,Yeh,Bendickson}
However, these approaches do not tell directly the optimal ARC parameters,
so the numerical calculation needs to be iterated
until these parameters are optimized. 
In this letter, we derive
analytic formulas of the structural parameters of an ARC that is applied
to a 1D PC. We deal with 1D systems for simplicity, but
the derived formulas would be applicable to multidimensional PCs
under appropriate approximation.

The structural parameters of a conventional ARC placed
between two homogeneous
media can be calculated easily if the refractive indices of the two media are
known.\cite{Yeh2,BornWolf,Macleod}
In Fig. \ref{fig:ARC} (a), the ARC (region 2) with
refractive index $n_2$ and thickness $d$ is placed between two
semi-infinite homogeneous media
with refractive indices $n_1$ (region 1) and $n_3$ (region 3).
These three regions are divided by two boundaries at $z=0$ and $-d$, where
the $z$ axis is defined as perpendicular to the surface of the ARC.
Each region consists of linear and lossless dielectrics.
The reflection coefficient at $z=-d$ is then given by
\begin{eqnarray}
 r = \frac{r_{1,2}+r_{2,3}{\rm exp}(2ik_2d)}
{1+r_{1,2}r_{2,3}{\rm exp}(2ik_2d)}\ ,
\label{eqn:ref_slab}
\end{eqnarray}
where $k_2$ is the normal component of the wave
vector in the ARC, and $r_{i,j}$ is the reflection coefficient of the light
propagating from region $i$ to $j$.\cite{Yeh2,BornWolf}
The reflection coefficient 
in Eq. (\ref{eqn:ref_slab})
equals zero for normal incidence light when $d=\lambda_0/4n_2$ 
and $n_2=\sqrt{n_1 n_3}$, where $\lambda_0$
is the wavelength in vacuum.\cite{Poisson}
%
\iftwocolumn
  \begin{figure}[\posfig]
   \begin{center}
    \leavevmode
    \epsfbox{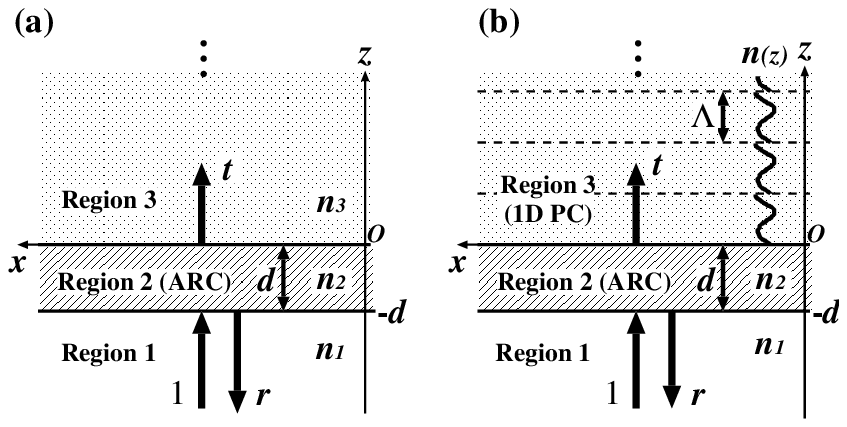}
    \caption{
    Antireflection coating (region 2) with thickness $d$ 
    and refractive index $n_2$ for (a) a semi-infinite homogeneous medium with 
    refractive index $n_3$ and for (b) a semi-infinite 1D PC with lattice constant 
    $\Lambda$ and periodic refractive index function $n(z) (=n(z+\Lambda))$. 
    }
    \label{fig:ARC}
   \end{center}
  \end{figure}
\fi

As in Fig. \ref{fig:ARC} (b), we replace the semi-infinite homogeneous
medium (region 3) in Fig. \ref{fig:ARC} (a) by a semi-infinite 1D PC
with lattice constant $\Lambda$ and 
periodic refractive index function $n(z)$. The function $n(z)$
has an arbitrary spatial modulation in the unit cell.
To determine the reflection coefficient for this case,
we expand the EM waves for the semi-infinite PC by all the eigenmodes of
the infinite PC.\cite{Minami}
These eigenmodes, which are extended into a complex $\Vec{k}$ space
(i.e., including the decaying waves),
in the infinite 1D PC can be obtained as the eigenvectors of a
transfer matrix\cite{Yeh,Bendickson,Minami,Pendry} with a periodic
boundary condition for a unit cell under a given frequency $\omega$,
a parallel component of a wave vector $\Vec{k}_\parallel$,
and a polarization of light $\sigma$.
The reflection coefficient $r$ of the electric field at the
interface $(z=-d)$ is determined by the continuity of 
the tangential components of EM waves at two boundaries ($z=0,-d$).
The expression of the derived reflection coefficient is
formally the same as Eq. (\ref{eqn:ref_slab}),  except
for $r_{2,3}$. 
The reflection coefficient $r_{2,3}$ for normal incidence light 
is modified into
\begin{eqnarray}
 r_{2,3}(\omega) &=& \frac{n_2-N(\omega)}{n_2+N(\omega)}\ ,
\label{eqn:ref23}
\end{eqnarray}
where $N(\omega)$ equals $\pm Y^{(\pm)}_{\Vec{k},\Vec{k}}(\omega)/\epsilon_0 c$,
which is a normalized surface admittance of the Bloch waves. 
Note that in $\pm Y^{(\pm)}_{\Vec{k},\Vec{k}}(\omega)$, the
``$+$''(``$-$'') sign applies to $Y^{(+)}_{\Vec{k},\Vec{k}}$
($Y^{(-)}_{\Vec{k},\Vec{k}}$).
The surface admittances $Y^{(\pm)}_{\Vec{k},\Vec{k}}$ for the
two orthogonal polarizations are defined as
\begin{eqnarray}
\label{eqn:admittance}
Y^{(+)}_{\Vec{k},\Vec{k}}= H_{\Vec{k},y}/E_{\Vec{k},x},\ 
Y^{(-)}_{\Vec{k},\Vec{k}}= H_{\Vec{k},x}/E_{\Vec{k},y},
\ {\rm at }\ z=+0, 
\end{eqnarray}
where $\Vec{k}$ is the Bloch wave vector, and  
$E_{\Vec{k},\xi}(H_{\Vec{k},\xi})$ with $\xi = x\ {\rm or}\ y$ stands 
for the tangential component of 
the electric(magnetic) field of the propagating Bloch waves with
a positive group velocity or the decaying Bloch waves 
in the positive $z$ direction.
These Bloch waves can be calculated from 
the forementioned transfer matrix, hence $N(\omega)$ in
Eq. (\ref{eqn:ref23}) is obtained directly. 
In the derivation of Eqs. (\ref{eqn:ref23}) and (\ref{eqn:admittance}) 
we used the fact that time reversal symmetry\cite{Jo2} inhibits
the simultaneous appearance of the propagating and the decaying modes
in the same direction for a given set $\{\omega$, $\Vec{k}_\parallel$  and
$\sigma\}$ in a semi-infinite 1D PC.
Note that the function $N(\omega)$ in Eq. (\ref{eqn:ref23})
is generally complex due to the phase difference 
between $E_{\Vec{k},\xi}$ and $H_{\Vec{k},\eta}$ 
($\xi,\eta = x\ {\rm or}\  y$)
of the Bloch waves at the interface, 
however the imaginary part of $N(\omega)$ is zero
if the surface of the semi-infinite PCs is 
a mirror plane in an infinite form of the PCs.
Note also that multiple reflections in the semi-infinite 1D PCs, 
which are represented as plane waves, are renormalized into 
a single Fresnel coefficient in Eq. (\ref{eqn:ref23}) by using 
the Bloch wave expansion.
%

The pair values of refractive index $n_2$ and thickness
$d$ of the ARC for semi-infinite 1D PCs that eliminate
reflectance are determined as follows.
The extremal condition of the reflectance\cite{BornWolf} 
$R (=|r|^2)$  with respect to $d$,
i.e., $\partial R/ \partial d = 0$,  is written as
\begin{eqnarray}
\label{eqn:D}
 D&\equiv& \frac{d}{\lambda_0/4n_2} = 
\frac{1}{2\pi i} {\rm ln }\left[\frac{r^*_{2,3}}{r_{2,3}}\right] 
+ m\ ,\ \   m=0,1,\cdots.
\end{eqnarray}
We call $D$ the ``normalized thickness''.
By substituting $d$ derived from Eq. (\ref{eqn:D}) into Eq. (\ref{eqn:ref_slab}), 
we obtain a refractive index $n_2$ that eliminates the reflectance:
\begin{eqnarray}
n_2 = \sqrt{n_1\RE(N)}
\sqrt{\frac{n_1-|N|^2/\RE(N)}{n_1-\RE(N)}}\ .
 \label{eqn:n2}
\end{eqnarray}
The necessary and sufficient condition for the perfect ARC for
semi-infinite 1D PCs
is $D\ge0$, $n_2\ge 1$ in Eqs. (\ref{eqn:D}) and  (\ref{eqn:n2}), and 
$\partial^2 R/\partial d^2 > 0$
under a given $n_1\ge 1$, $N(\omega)$, and $\omega$. 
\iftwocolumn
\begin{figure}[\posfig]
 \begin{center}
  \leavevmode
  \epsfbox{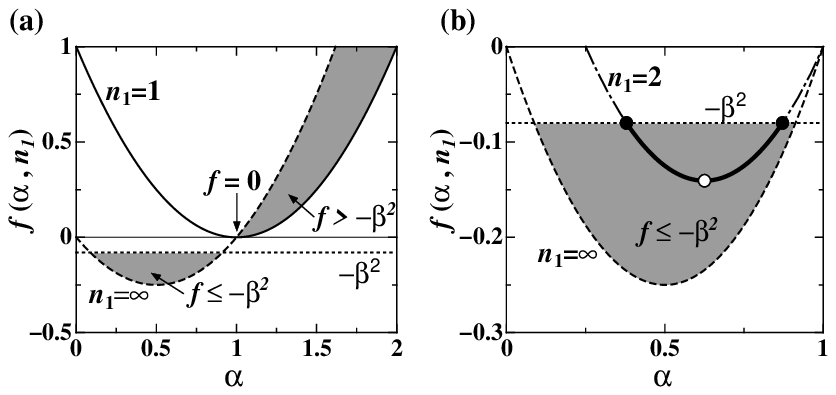}
  \caption{(a) Illustration of sufficient conditions in Eq. (6). 
  The shaded areas indicate the region satisfying the conditions 
  with $-\beta^2=-0.08$ (dotted line) as an example. (b) Magnification 
  of region $0< \alpha < 1$ and $-0.3\le f \le0$ of Fig. 2(a). 
  Function $f$ with $n_1=2$ is also plotted.}
  \label{fig:f}
 \end{center}
\end{figure}
\fi
%

Next, we investigate
the physical conditions encapsulated in Eq. (\ref{eqn:n2}).
The condition $\RE(N)>0$ is necessary
to obtain a positive real number for $n_2$, so the explicit form of $n_2\ge 1$ 
with Eq.(\ref{eqn:n2}) is classified by
using positive $\alpha \equiv \RE(N)/n_1$:
\begin{eqnarray}
\label{eqn:n2cond}
 n_2\ge1 \Leftrightarrow 
 \left\{
  \begin{array}{ll}
   {\rm (A)}\  f >   -\beta^2 &  \mbox{ for}\ \ 1<\alpha<\infty \\
   {\rm (B)}\  f =   -\beta^2 &  \mbox{ for}\ \  \alpha=1 \ \ \ \ \ \ \ \ \ ,\\
   {\rm (C)}\  f \le -\beta^2 &  \mbox{ for}\ \  0<\alpha<1 
  \end{array}
 \right.
\end{eqnarray}
where $f(\alpha,n_1)\equiv(\alpha-1)(\alpha- 1/n_1^2)$,
and $\beta \equiv \IM(N)/n_1$.

These sufficient conditions in the rhs of Eq. (\ref{eqn:n2cond}) are 
illustrated in Fig. \ref{fig:f} (a). 
As $n_1$ changes from 1 to $\infty$,
the curve for $f$ as a function of $\alpha$ varies
from a solid one to a broken one continuously. 
The shaded area indicates
the region satisfying the conditions of Eq. (\ref{eqn:n2cond}).
When $1<\alpha<\infty$, $f>-\beta^2$ always holds because $f>0$. 
The case $\alpha=1$ leads to $\beta=0$ due to $f=0$.
A magnification of region $0<\alpha<1$ 
in Fig. \ref{fig:f} (a) is shown in Fig. \ref{fig:f} (b),
where a curve for function $f$ with $n_1=2$ is added.
The thick solid line shows the values 
of $f$ and $\alpha$ that satisfy the third condition in the rhs
of Eq. (\ref{eqn:n2cond}) for $n_1=2$ and a given $-\beta^2$.
This curve connects two crossing points (filled circles) of $f$ and $-\beta^2$ 
through the minimum point (open circle) of $f$.
These three points are used to rewrite 
the inequality $f\le-\beta^2$ for $0<\alpha<1$
in Eq. (\ref{eqn:n2cond}),  which leads to two conditions: 
(1) the minimum point of $f$ is less than or equal to $-\beta^2$,
and (2) $\alpha$ is located between the two crossing points.

By applying the previous,
we can obtain the final form of the three sufficient conditions for
$n_2\ge1$ in Eq. (\ref{eqn:n2cond}):
\begin{eqnarray}
\label{eqn:A}
& {\rm (A)}\ \ &  1< \alpha < \infty\ , \\
\label{eqn:B}
& {\rm (B)}\ \ &  \alpha = 1\ {\rm and}\ \beta=0\ ,  \\
& {\rm (C)}\ \ &  0 < \alpha < 1\ ,\   0 \le g\  {\rm and}\ \nonumber \\
\label{eqn:C}
&& -\sqrt{g} \le  1+1/n_1^2-2\alpha  \le \sqrt{g}\ . 
\end{eqnarray}
where $g(\beta,n_1)\equiv \left(1-1/n_1^2\right)^2 - 4\beta^2$.
Note that the condition (C) includes $n_1\neq 1$ implicitly.
Conditions (A)-(C) depend on $n_1$ and $N(\omega)$ only, 
so the $n_2$ needed to achieve
ARCs for semi-infinite 1D PCs can be directly obtained by using 
Eqs. (\ref{eqn:A})-(\ref{eqn:C}) and (\ref{eqn:n2}).
Accordingly, thickness $d$ is also determined by Eq. (\ref{eqn:D}) 
and $\partial^2 R/\partial d^2 > 0$.
\iftwocolumn
\begin{figure}[\posfig]
 \begin{center}
  \leavevmode
  \epsfbox{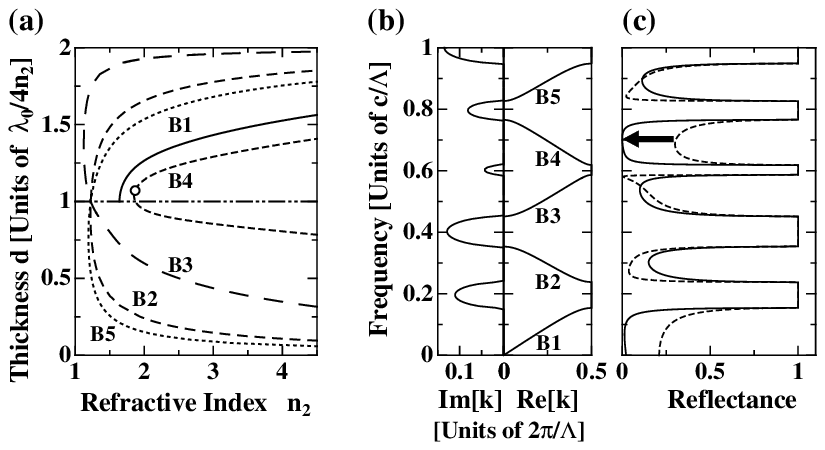}
  \caption{(a) Pairs of \{$n_2$ and $d$\} needed to achieve ARC for
  a semi-infinite 1D PC. 
  (b) Dispersion relation of propagating and decaying modes. 
  (c) Reflectance with (solid line) and without (broken line) ARC.}
  \label{fig:opt}
 \end{center}
\end{figure}
\fi
%

To give a practical example of ARCs for semi-infinite 1D PCs,
in Fig. {\ref{fig:opt}}(a) we plot the refractive
index and the thickness needed to form the ARC for a semi-infinite 1D PC.
We assume that $n_1=1$, that the unit cell of the PC consists of 
Si ($n=3.5$) and ${\rm SiO}_2$ ($n=1.5$) with the same thickness
$\Lambda/2$,  and that the  ${\rm Si}$ layer 
is the boundary layer between the semi-infinite PC and the ARC.
In Fig. {\ref{fig:opt}}(b), the dispersion relation of the infinite 1D PC
is plotted and the band index is indicated by B1-B5.
When the frequency changes along a band, 
the corresponding pair of $[n_2(\omega)$ and $d(\omega)]$ 
forms a continuous trajectory [as shown in Fig. \ref{fig:opt}(a)].
Note that in Fig. \ref{fig:opt}(a), only condition (A) is needed 
to calculate $n_2$ and $d$, and
frequency regions near the band edges that require $n_2 \ge 4.5$ 
are not plotted. 
Note also that the deviation in the normalized thickness from 1
is due to the phase shift  
between the electric and the magnetic fields at the interface $(z=+0)$.

In Fig. {\ref{fig:opt}}(c) we show the reflectance of the 
1D PC with (solid line) and without (broken line) an ARC.
The ARC was designed at a selected frequency ($\omega\Lambda/2\pi c=0.7$), 
which is indicated by an arrow, 
where $N(=3.384 -{\rm i}0.499)$ fulfills condition (A).
The required values of $n_2 (=1.868)$ and $D (=1.071)$ for the ARC
are calculated from Eqs. (\ref{eqn:n2}) and (\ref{eqn:D}) with $m=1$,
which is shown by a circle in Fig. {\ref{fig:opt}}(a).
In Fig. \ref{fig:opt}(c),
the ARC eliminates the reflectance (solid line)
at the selected frequency (arrow).
%

In conclusion, we have presented a systematic method for 
designing a perfect ARC for a semi-infinite 1D PC.
We derived exact formulas of structural parameters of the ARC and 
Fresnel coefficient of the PC.
The concept of $\Vec{k}$ matching\cite{Mekis2,Mekis3} was not used
for our derivations, and the Bloch wave vector in the infinite PC 
is not included explicitly in Eqs. (\ref{eqn:ref23}) 
and (\ref{eqn:admittance}).
The surface immittance (admittance and impedance) of the PC
plays an essential role in the formulas.
The adiabatic interconnection\cite{Baba,Xu,Mekis2} can be
interpreted to be gradual immittance-matching structures.
Under appropriate approximation, 
the designing method would to a large extent be applicable
to multidimensional PCs that have a position-dependent 
surface-immittance across the surface.
%

The authors thank
Professor K. Cho for fruitful discussions and 
for showing us their work (see Ref. \onlinecite{Minami}) before publication,
Y. Miyamoto, O. Sugino, Professor K. Ohtaka, K. Hirose, A. Gomyo, 
R. Kuribayashi, M. Saito, Professor H. Ishihara,  T. Ikawa, T. Hirai,
T. Minami, H. Ajiki, K. Suzuki, and Y. Watanabe for fruitful discussions, 
and K. Nakamura and J. Sone for their support.
This work was supported in part by Special Coordination Funds for
Promoting Science and Technology.
J. Ushida
dedicates this work to Professor and Mrs. Cho in honor 
of their 60th birthdays.
\iftwocolumn
%
\fi

%
\iftwocolumn
\else
\newpage
\begin{large}
\underline{\bf List of Figure Captions}\\
\end{large}

\noindent
Fig. 1: Antireflection coating (Region 2) with thickness $d$ 
and refractive index $n_2$ for (a) a semi-infinite homogeneous medium with 
refractive index $n_3$ and for (b) a semi-infinite 1D PC with lattice constant 
$\Lambda$ and periodic refractive index function $n(z) (=n(z+\Lambda))$. \\

\noindent
Fig. 2: (a) Illustration of sufficient conditions in Eq. (6). 
The shaded areas indicate the region satisfying the conditions 
with $-\beta^2=-0.08$ (dotted line) as an example. (b) Magnification 
of region $0< \alpha < 1$ and $-0.3\le f \le0$ of Fig. 2(a). 
Function $f$ with $n_1=2$ is also plotted.\\

\noindent
Fig. 3: (a) Pairs of \{$n_2$ and $d$\} needed to achieve ARC for
a semi-infinite 1D PC. 
(b) Dispersion relation of propagating and decaying modes. 
(c) Reflectance with (solid line) and without (broken line) ARC.\\

%
\newpage
\begin{figure}[\posfig]
 \begin{center}
  \leavevmode
  \epsfbox{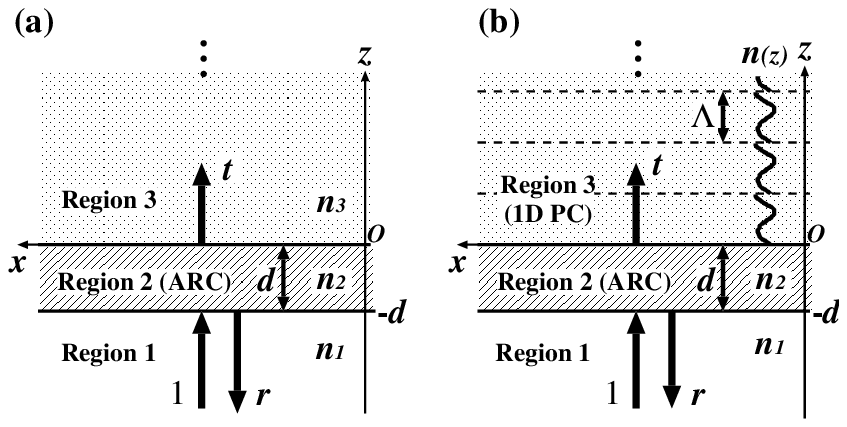}
  \begin{center}
   \vspace*{0.5cm}
   
   Jun Ushida, Masatoshi Tokushima, Masayuki Shirane, and Hirohito Yamada
  \end{center}
  \caption{}
  \label{fig:ARC}
 \end{center}
\end{figure}
\newpage
\begin{figure}[\posfig]
 \begin{center}
  \leavevmode
  \epsfbox{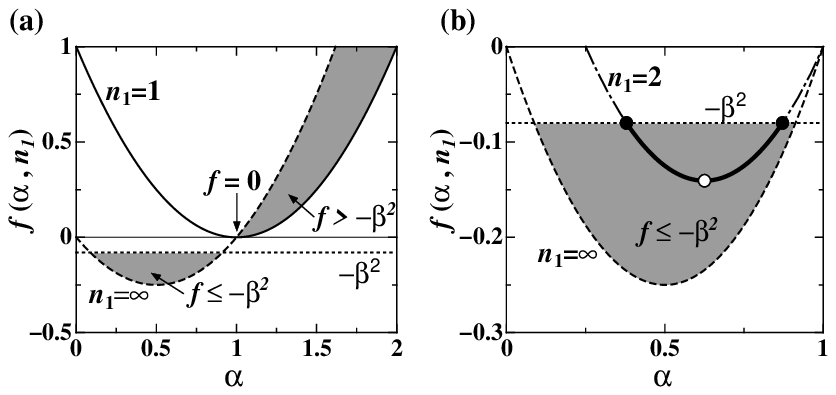}
  \begin{center}
   \vspace*{0.5cm}
   
   Jun Ushida, Masatoshi Tokushima, Masayuki Shirane, and Hirohito Yamada
  \end{center}
  \caption{}
  \label{fig:f}
 \end{center}
\end{figure}
%
\newpage
\begin{figure}[\posfig]
 \begin{center}
  \leavevmode
  \epsfbox{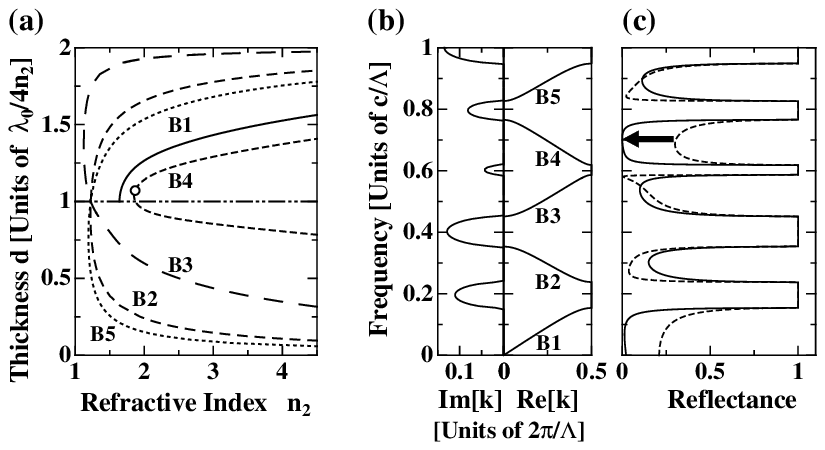}
  \begin{center}
   \vspace*{0.5cm}
   
   Jun Ushida, Masatoshi Tokushima, Masayuki Shirane, and Hirohito Yamada
  \end{center}
  \caption{}
  \label{fig:opt}
 \end{center}
\end{figure}
\fi
\end{document}